\documentclass[10pt]{article}
\usepackage{graphicx}
\usepackage{amsmath}
\usepackage{amssymb}
\usepackage{caption2}
\begin{document}
\bibliographystyle{prsty}
\begin{center}
{\large {\bf \sc{Analysis of the strong decay $X(4140)\rightarrow J/\psi \phi$  via the light-cone QCD sum rules }}} \\[2mm]
Zun-Yan Di \footnote{E-mail: dizunyan@126.com.  }, Zhi-Gang Wang \footnote{E-mail: zgwang@aliyun.com.  }    \\
Department of Physics, North China Electric Power University, Baoding 071003, P. R. China
\end{center}

\begin{abstract}
In this article, we take the $X(4140)$ as the axialvector tetraquark state with the symbolic quark structure $[sc]_S[\bar{s}\bar{c}]_A+[sc]_A[\bar{s}\bar{c}]_S$,
and calculate the width of the two-body strong decay
$X(4140)\rightarrow J/\psi \phi$
within the framework of the light-cone sum rules. Different from the traditional light-cone sum rules, at the phenomenological side, we introduce  parameters $C$ to eliminate the contaminations from the higher resonances and continuum states, and match the hadron side with the QCD side of the correlation function based
on rigorous  quark-hadron duality to obtain the stable QCD sum rules.
Then we obtain the decay width $\Gamma(X(4140)\rightarrow J/\psi \phi)=145\pm21\, \text{MeV}$,
which is reasonable according to the experimental data $162\pm21^{+24}_{-49 } \,\text{MeV}$ from the LHCb collaboration.
The numerical result supports the possibility that the $X(4140)$ could be
the $[sc]_S[\bar{s}\bar{c}]_A+[sc]_A[\bar{s}\bar{c}]_S$ type axialvector tetraquark state.
\end{abstract}

PACS number: 12.39.Mk, 14.20.Lq, 12.38.Lg

Key words: Tetraquark state, Energy scale, QCD sum rules

\section{Introduction}
In 2009,
the CDF collaboration observed the $X(4140)$  in the $J/\psi \phi$ mass spectrum in the exclusive $B^+ \rightarrow J/\psi \phi K^+$  decay for the first time with a statistical significance more than $3.8\sigma$ \cite{CDF:2009jgo}.
Then the $X(4140)$ was confirmed by CDF, CMS, D0, LHCb collaborations \cite{CDF:2009jgo,CDF:2011pep,CMS:2013jru,D0:2013jvp,D0:2015nxw,LHCb:2016axx,LHCb:2016nsl}.
In 2016, the LHCb collaboration performed the first full amplitude analysis of the  $B^+ \rightarrow J/\psi \phi K^+$ decay
and confirmed the  $X(4140)$ and $X(4274)$ in the $J/\psi \phi$  mass spectrum with statistical significances $8.4\sigma$ and $6.0\sigma$ , respectively,
and determined the quantum numbers  to be $J^{PC}=1^{++}$  with statistical significances $5.7\sigma$ and $5.8\sigma$, respectively \cite{LHCb:2016axx,LHCb:2016nsl}. Furthermore, the LHCb collaboration observed  the $X(4500)$ and $X(4700)$ in the $J/\psi \phi$  mass spectrum  with statistical significances $6.1\sigma$ and $5.6\sigma$, respectively,  and determined the quantum numbers $J^{PC} =0^{++}$ with statistical significances $4.0\sigma$ and $4.5\sigma$, respectively \cite{LHCb:2016axx,LHCb:2016nsl}.
In 2021,
the LHCb collaboration performed an improved full amplitude analysis of the $B^+ \rightarrow J/\psi \phi K^+$ decay using $pp$ collision data corresponding to a total integrated luminosity of $9 \, \text{fb}^{-1}$.
The four $X$ structures, i.e. $X(4140)$, $X(4274)$, $X(4500)$ and $X(4700)$, which decaying to the final states $J/\psi \phi$,
were confirmed with higher significances.
Notably, the width  $\Gamma_{X(4140)}=162 \pm 21_{-49}^{+24} \, \text{MeV} $
was substantially larger than that  determined previously \cite{LHCb:2021uow}.

Since the $X(4140)$ was observed in the $J/\psi \phi$ invariant mass spectrum,
and the $J/\psi \phi$ system contains a $c \bar{c}$ pair and a $s \bar{s}$ pair,
if the $X(4140)$'s dominant Fock components are four-quark states, irrespective of the color singlet-singlet type or antitriplet-triplet type,
the valence quark constituents are symbolically $c \bar{c} s \bar{s}$.
The $D_s^* \bar{D}_s^*$ threshold is 4224.4MeV from the Particle Data Group \cite{ParticleDataGroup:2024cfk},
which leads to the possible molecule assignment for the $X(4140)$.
The LHCb collaboration established the quantum numbers of the $X(4140)$ as $J^{PC} = 1^{++}$,
which ruled out the overwhelming $D_s^* \bar{D}_s^*$ molecule identification with the assumption of the quantum numbers $J^{PC} = 0^{++}$ or $2^{++}$
after the discovery of the X(4140),
but did not rule out the existence of the $D_s^* \bar{D}_s^*$ molecular states with the $J^{PC} = 0^{++}$ and $2^{++}$,
those molecular states might be observed experimentally in the future.

The possible assignments for the $X(4140)$ are tetraquark state \cite{Stancu:2009ka,Wang:2015pea,Lebed:2016yvr,
Maiani:2016wlq,Zhu:2016arf,Lu:2016cwr,Wu:2016gas,Anwar:2018sol,
Wang:2018qpe,Chen:2010ze,Agaev:2017foq,Wang:2021ghk,Wang:2024ciy,Wang:2025sic},
hybrid state \cite{Mahajan:2009pj,Wang:2009ue,Wang:2009ry} or rescattering effect \cite{Liu:2016onn}, etc.
In this work,
we focus on the identifications of the $X(4140)$ as a tetraquark state.
In Ref.\cite{Lebed:2016yvr}, Lebed and Polosa identified the $X(3915)$ as the 1S $[cs]_S [\bar{c} \bar{s}]_S$ tetraquark state with
the $J^{PC} = 0^{++}$ due to lacking of decaying to the meson pairs $D \bar{D}$ and $D^* \bar{D}^*$, and attributed the only known decay to the  $J/\psi \omega$ to the $ \omega - \phi$ mixing effects,
and identified the $X(4140)$ as the 1S $[cs]_S[\bar{c}\bar{s}]_A+[cs]_A[\bar{c}\bar{s}]_S$ tetraquark state with the $J^{PC} = 1^{++}$ based on the effective spin-spin and spin-orbit Hamiltonian, the $X(3915)$ and $X(4140)$ are cousins.
Then in Ref.\cite{Maiani:2016wlq}, Maiani, Polosa and Riquer took the mass of the $X(4140)$ as input parameter,
and obtained the mass spectrum of the diquark-antidiquark type $ c \bar{c} s \bar{s}$ tetraquark states with positive parity,
however, there was no room to accommodate  the $X(4274)$.

On the other hand, in Ref.\cite{Stancu:2009ka}, F. Stancu studied   the $c \bar{c} s \bar{s}$ tetraquark  mass spectrum  via a
simple quark model with the chromomagnetic interaction but no correlated quarks,
and obtained two lowest masses $4195\, \rm{MeV}$ and $4356\,\rm{MeV}$ with the $J^{PC} = 1^{++}$. Again, there is no room to accommodate  the $X(4274)$.

The QCD sum rules is a powerful theoretical method in studying  the exotic states \cite{Wang:2025sic,WangZG-landau-PRD,Nielsen-JPG-Review}. In Refs.\cite{Chen:2010ze,Agaev:2017foq}, the authors accomplished   the operator product expansion up to the vacuum condensates of
dimension 8 to study the $[sc]_S[\bar{s}\bar{c}]_A+[sc]_A[\bar{s}\bar{c}]_S$ tetraquark state with the $J^{PC}= 1^{++}$,
Chen and Zhu obtained the  mass $4.07\pm0.10\, \rm{GeV}$  \cite{Chen:2010ze},
Agaev, Azizi, and Sundu obtained the  mass $4.183\pm0.115 \, \rm{GeV}$  \cite{Agaev:2017foq},
which were all compatible with the experimental data $4146.5\pm3.0$MeV from the Particle Data Group within uncertainties \cite{ParticleDataGroup:2024cfk},
however, the pole contributions are $44.4\%$ and $23\,\%$, respectively.

In Ref.\cite{Wang:2024ciy}, we took account of the light-flavor $SU(3)$ breaking effects comprehensively to study the $c\bar{c}s\bar{s}$ mass spectrum,
and revisited the assignments of the potential $cs \bar{c} \bar{s}$ tetraquark candidates
and superseded the old assignments \cite{Wang:2018qpe,Wang:2021ghk,Wang:2016ujn,Wang:2016gxp}.
The $X(4140)$ was identified as the 1S $[sc]_S[\bar{s}\bar{c}]_A+[sc]_A[\bar{s}\bar{c}]_S$ or $[sc]_S[\bar{s}\bar{c}]_{\tilde{A}}+[sc]_{\tilde{A}}[\bar{s}\bar{c}]_S$
tetraquark state with the quantum numbers $J^{PC} = 1^{++}$.

In this article, we take the $X(4140)$ as the 1S $[sc]_S[\bar{s}\bar{c}]_A+[sc]_A[\bar{s}\bar{c}]_S$ tetraquark state with the $J^{PC} = 1^{++}$,
and choose  the interpolating current from Ref.\cite{Wang:2024ciy} to study the two-body strong decay $X(4140)\rightarrow J/\psi \phi$ via the light-cone sum rules.
Different from the traditional light-cone sum rules,
at the phenomenological side, we introduce  parameters $C$
to eliminate the contaminations from the higher resonances and continuum states,
and match the hadron side with the QCD side of the correlation function
based on rigorous quark-hadron duality to obtain the stable QCD sum rules with the variations of the Borel parameters \cite{Wang:2025sic}.
The rigorous quark-hadron duality for the three-point correlation functions, which worked very well,  was suggested  in our previous works \cite{Wang:2025sic}.
For the details of the rigorous quark-hadron duality, one can consult Refs.\cite{Wang:2025sic,Wang:2017lot,Wang:2019iaa}.

This article is organized as follows: in section 2, we calculate the width of the $X(4140)$ as the $J^{PC}=1^{++}$ tetraquark state via the light-cone sum rules, and present the full technical details; we perform numerical analysis in section 3;
the last section is reserved for our conclusion.

\section{Light-cone sum rules for the $X(4140)$ }
In the light-cone sum rules, we study the two-body strong decay $X(4140)\rightarrow J/\psi \phi$ with the two-point correlation function,
\begin{eqnarray}\label{a1}
\Pi_{\alpha \beta}(p,q)&=&i\int d^{4} x e^{iq\cdot x} \langle \phi(p)|T\left\{J_{\alpha}^{J/\psi}(x) J_{\beta}^{X\dag}(0)\right\}|0\rangle \ ,
\end{eqnarray}
where the $p$ is the four-momentum of the $\phi$ meson.
The currents $J_{\alpha}^{J/\psi}(x)$ and $J_{\beta}^{X}(x)$  are the interpolating currents of the mesons $J/\psi$ and $X(4140)$, respectively \cite{Wang:2024ciy},
\begin{eqnarray}\label{a2}
J_{\alpha}^{J/\psi}(x)&=& \bar{c}(x)\gamma_{\alpha}c(x)\ , \nonumber\\
J_{\beta}^{X}(x)&=&\frac{\varepsilon^{i j k}\varepsilon^{i m n}}{\sqrt{2}}
\left[ s^{T}_j(x)C\gamma_{5}c_{k}(x) \bar{s}_{m}(x) \gamma_{\beta}C \bar{c}_n^{ T}(x) +s^{ T}_j (x)C\gamma_{\beta}c_{k}(x) \bar{s}_{m}(x) \gamma_5C \bar{c}_n^{ T}(x)\right]\, ,
\end{eqnarray}
where the  $i$, $j$, $k$, $m$ and $n$ are color indexes,
the $C$ is the charge conjugation matrix, and the $c$ and $s$ represent the charm and strange quarks, respectively.

At the phenomenological side, we insert a complete set of intermediate hadronic states with the same quantum numbers as the current operators
$J_{\alpha}^{J/\psi}(x)$ and $J_{\beta}^{X}(x)$
into the two-point correlation function $\Pi_{\alpha \beta}(p,q)$,
and isolate the ground state contributions to obtain the result,
\begin{eqnarray}\label{a22}
\Pi_{\alpha\beta}(p,q)&=&\frac{\lambda_{X}m_{J/\psi}f_{J/\psi}g_{XJ/\psi\phi}\varepsilon^{\lambda \tau \rho \theta}p^\prime_\lambda \xi^*_\rho \zeta_\alpha \zeta^*_\theta \varepsilon_\tau \varepsilon^*_\beta}{(m^2_{J/\psi}-q^2)(m^2_X-p^{\prime2})}+\cdots\nonumber\\
&=&\frac{\lambda_{X}m_{J/\psi}f_{J/\psi}g_{XJ/\psi\phi}\varepsilon^{\lambda \tau \rho \theta}p^\prime_\lambda \xi^*_\rho }{(m^2_{J/\psi}-q^2)(m^2_X-p^{\prime2})}\left (-g_{\alpha\theta}+\frac{q_\alpha q_\theta}{q^2}\right)\left(-g_{\beta\tau}+\frac{p^\prime_\beta p^\prime_\tau}{p^{\prime2}}\right)+\cdots \nonumber\\
&=&\bigg[\frac{\lambda_{X}m_{J/\psi}f_{J/\psi}g_{XJ/\psi\phi}}{
(m^2_{J/\psi}-q^2)(m^2_X-p^{\prime2})}+\frac{1}{m^2_X-p^{\prime2}}
\int^{\infty}_{s^0_{J/\psi}}dt\frac{\rho_{\psi^\prime}(p^{\prime2},t)}{t-q^2} \nonumber\\
&&+\frac{1}{m^2_{J/\psi}-q^2}\int^{\infty}_{s^0_{X}}dt
\frac{\rho_{X^\prime}(q^2,t)}{t-p^{\prime2}}+\cdots \bigg] \nonumber\\
&&\times \varepsilon^{\lambda \tau \rho \theta}p^\prime_\lambda \xi^*_\rho  \left (-g_{\alpha\theta}+\frac{q_\alpha q_\theta}{q^2}\right)\left(-g_{\beta\tau}+\frac{p^\prime_\beta p^\prime_\tau}{p^{\prime2}}\right)+\cdots\, ,
\end{eqnarray}
where the $g_{XJ/\psi\phi}$ is the hadronic coupling constant defined by
\begin{eqnarray}\label{a23}
\langle\phi(p,\xi)J/\psi(q,\zeta)| X(p',\varepsilon) \rangle&=&i g_{XJ/\psi\phi} \varepsilon^{\lambda \tau \rho \theta}p'_\lambda \varepsilon_\tau  \xi^*_\rho  \zeta^*_\theta \ ,
\end{eqnarray}
the decay constants $f_{J/\psi}$ and $\lambda_X$ are defined by
\begin{eqnarray}\label{a24}
\langle 0|J ^{J/\psi}_\alpha (0) | J/\psi(q,\zeta)\rangle&=&m_{J/\psi} f_{J/\psi} \zeta_\alpha\, ,
\end{eqnarray}
\begin{eqnarray}\label{a25}
\langle X(p',\varepsilon)|J^{X\dag}_\beta (0) | 0\rangle&=&\lambda_X \varepsilon_\beta^* \, ,
\end{eqnarray}
the $\xi_\rho$, $\zeta_\theta$ and $\varepsilon_\beta$
are the polarization vectors of the $\phi$, $J/\psi$ and $X(4140)$ respectively,
the $s^0_{J/\psi}$ and $s^0_X$ are the continuum threshold parameters,
the two functions $\rho_{\psi^\prime}(p'^2,t)$ and $\rho_{X^\prime}(q^2,t)$
have complex dependence on the transitions between the ground states and the higher resonances or the continuum states.

We introduce the parameters $C_{\psi'}$ and $C_{X'}$ to parameterize the net effects,
\begin{eqnarray}\label{a26}
C_{\psi'}&=&\int^{\infty}_{s^0_{J/\psi}}dt\frac{\rho_{\psi'}(p'^2,t)}{t-q^2}\ ,
\end{eqnarray}
\begin{eqnarray}\label{a27}
C_{X'}&=&\int^{\infty}_{s^0_{X}}dt\frac{\rho_{X'}(q^2,t)}{t-p'^2}\ .
\end{eqnarray}
Then the correlation function $\Pi_{\alpha \beta}(p,q)$
on the phenomenological side can be written as
\begin{eqnarray}\label{a28}
\Pi_{\alpha\beta}(p,q)
&=&\left[\frac{\lambda_{X}m_{J/\psi}f_{J/\psi}g_{XJ/\psi\phi}}{
(m^2_{J/\psi}-q^2)(m^2_X-p'^2)}+\frac{C_{\psi'}}{m^2_X-p'^2}
+\frac{C_{X'}}{m^2_{J/\psi}-q^2}+\cdots \right]\, , \nonumber\\
&&\times \varepsilon^{\lambda \tau \rho \theta}p'_\lambda \xi^*_\rho  \left (-g_{\alpha\theta}+\frac{q_\alpha q_\theta}{q^2}\right)\left(-g_{\beta\tau}+\frac{p'_\beta p'_\tau}{p'^2}\right)+\cdots\, .
\end{eqnarray}

The tensor structures in the correlation function $\Pi_{\alpha\beta}(p,q)$ are complex, we should simplify them, and thus facilitate the calculations at the QCD side.
We project out the component $\Pi(p'^2,q^2)$ by introducing the operator $P^{\alpha\beta}$,
\begin{eqnarray}\label{a29}
\Pi(p'^2,q^2)\,V\cdot\xi^*&=&P^{\alpha\beta}\Pi_{\alpha\beta}(p,q)\nonumber\\
&=&\left[\frac{\lambda_{X}m_{J/\psi}f_{J/\psi}g_{XJ/\psi\phi}}{
(m^2_{J/\psi}-q^2)(m^2_X-p'^2)}+\frac{C_{\psi'}}{m^2_X-p'^2}
+\frac{C_{X'}}{m^2_{J/\psi}-q^2}+\cdots \right] \nonumber\\
&&\times \left(m^2_X+m^2_{J/\psi}-m^2_\phi \right)V\cdot\xi^*+\cdots\ ,
\end{eqnarray}
where $P^{\alpha\beta}=V_{\tau'}q_{\lambda'}\varepsilon^{\lambda'\tau'\beta\alpha}$,
the unit vector $V_\tau$ is defined by us, which satisfies the conditions: $V^2=1$, $V\cdot q=0$ and $V\cdot p=0$, but it does not have a specific expression.
Thereafter, we would like to add a subscript $H$ to represent  the hadron side, i.e. $\Pi_H(p'^2,q^2)$.

Now we briefly outline the operator product expansion for the correlation function $\Pi_{\alpha \beta}(p,q)$.
We contract the quark fields $s$ and $c$ in the correlation function $\Pi_{\alpha \beta}(p,q)$
with Wick theorem, and obtain the result,
\begin{eqnarray}\label{a211}
\Pi_{\alpha \beta}(p,q)&=&\frac{i \varepsilon^{i j k} \varepsilon^{i m n}}{\sqrt{2}}\int d^{4}x e^{iq\cdot x} \nonumber\\
&&\bigg\{\frac{1}{2}\langle\phi(p)|\bar{s}_j(0)s_m(0)|0\rangle Tr\left[\gamma_\beta \gamma_5 C S_{a k}^{T}(x)C \gamma_\alpha C S_{n a}^{ T}(-x)C\right]  \nonumber\\
&&+\frac{1}{4} \langle\phi(p)|\bar{s}_j(0) \gamma^\lambda s_m(0)|0\rangle
\bigg [ Tr\left[\gamma_\beta \gamma_\lambda \gamma_5 C S_{a k}^{ T}(x)C \gamma_\alpha C S_{n a}^{ T}(-x)C \right]   \nonumber\\
&&- Tr\left[ \gamma_5 \gamma_\lambda \gamma_\beta C S_{a k}^{ T}(x)C \gamma_\alpha C S_{n a}^{ T}(-x)C \right]  \bigg ]  \nonumber\\
&&+\frac{1}{8} \langle\phi(p)|\bar{s}_j(0) \sigma^{\lambda \tau} s_m(0)|0\rangle
\bigg [ Tr\left[\gamma_\beta \sigma_{\lambda \tau} \gamma_5 C S_{a k}^{ T}(x)C \gamma_\alpha C S_{n a}^{ T}(-x)C \right]   \nonumber\\
&&- Tr\left[ \gamma_5 \sigma_{\lambda \tau} \gamma_\beta C S_{a k}^{ T}(x)C \gamma_\alpha C S_{n a}^{ T}(-x)C \right]  \bigg ] \nonumber\\
&&-\frac{1}{4} \langle\phi(p)|\bar{s}_j(0) \gamma^\lambda \gamma_5 s_m(0)|0\rangle
\bigg [ Tr\left[\gamma_\beta \gamma_\lambda  C S_{a k}^{ T}(x)C \gamma_\alpha C S_{n a}^{ T}(-x)C \right]  \nonumber\\
&&+ Tr\left[  \gamma_\lambda \gamma_\beta C S_{a k}^{ T}(x)C \gamma_\alpha C S_{n a}^{ T}(-x)C \right]  \bigg ] \bigg\} \ ,
\end{eqnarray}
where the $S_{ij}(x)$ is the full $c$ quark propagator,
\begin{eqnarray}
S_{i j}(x)&=&\frac{i}{(2\pi)^4}\int d^4 ke^{-ik\cdot x}\bigg\{\frac{k\!\!\!/ +m_{c}}{k^{2}-m_{c}^{2}}\delta_{i j}-g_{s}t_{i j}^{n}G_{\alpha\beta}^{n}\frac{(k\!\!\!/+m_{c})\sigma^{\alpha\beta}+\sigma^{\alpha\beta}(k\!\!\!/+m_{c})}{4(k^{2}-m_{c}^{2})^{2}} +\cdots\bigg\} \ ,
\end{eqnarray}
and $t^n=\frac{\lambda^n}{2}$, the $\lambda^n$ is the Gell-Mann matrix.

The matrix elements of the vacuum-$\phi(p)$  in the correlation function $\Pi_{\alpha \beta}(p,q)$,
which can be expanded in terms of the corresponding light-cone distribution amplitudes at zero point \cite{Xie:2022ilz,Ball:2007zt,Ball:1996tb,Ball:2007rt},
\begin{eqnarray}
\langle\phi(p,\xi)|\bar{s}(0) \gamma_\lambda s(0)|0\rangle&=&f_\phi^{\|}m_\phi \xi_{\lambda}^{ *} \, ,
\end{eqnarray}
\begin{eqnarray}
\langle\phi(p,\xi)|\bar{s}(0) \sigma_{\lambda \tau} s(0)|0\rangle&=&if_\phi^{\bot}( \xi_{\lambda}^{ *}p_\tau-\xi_{\tau}^{ *}p_\lambda) \, ,
\end{eqnarray}
\begin{eqnarray}
\langle\phi(p,\xi)|\bar{s}(0) t^{n}G_{\alpha' \beta'}^{n} s(0)|0\rangle&=&-if_\phi^{\bot}m_\phi^2\zeta_{4\phi}^\bot \left( \xi_{\alpha'}p_{\beta'}-\xi_{\beta'}p_{\alpha'}\right) \ ,
\end{eqnarray}
\begin{eqnarray}
\langle\phi(p,\xi)|\bar{s}(0) \gamma_\lambda t^{n}G_{\alpha' \beta'}^{n} s(0)|0\rangle&=&if_\phi^{\|}m_\phi \kappa_{3\phi}^\|
\bigg [\xi_{\alpha'}\left( p_{\beta'}p_{\lambda}-\frac{1}{3}m_\phi^2 g_{\beta' \lambda}\right)
-\xi_{\beta'}\left( p_{\alpha'}p_{\lambda}-\frac{1}{3}m_\phi^2 g_{\alpha' \lambda}\right) \bigg] \nonumber\\
&&+\frac{i}{3}f_\phi^{\|}m_\phi^3 \kappa_{4\phi}^\| \left(\xi_{\alpha'}g_{\beta' \lambda}-\xi_{\beta'}g_{\alpha' \lambda}\right)  \ ,
\end{eqnarray}
\begin{eqnarray}
\langle\phi(p,\xi)|\bar{s}(0) \gamma^\lambda \gamma_5 t^{n}G_{\alpha' \beta'}^{n} s(0)|0\rangle
&=&-\frac{1}{2}\varepsilon_{\mu\nu \alpha' \beta'}
\bigg \{ f_\phi^{\|}m_\phi \zeta_{3\phi}^\|
\bigg [\xi^{\mu} \left( p^{\nu}p^{\lambda}-\frac{1}{3}m_\phi^2 g^{\nu \lambda}\right) \nonumber\\
&&-\xi^{\nu}\left( p^{\mu}p^{\lambda}-\frac{1}{3}m_\phi^2 g^{\mu \lambda}\right) \bigg] \nonumber\\
&&+\frac{1}{3}f_\phi^{\|}m_\phi^3 \zeta_{4\phi}^\| \left(\xi^{\mu}g^{\nu \lambda}-\xi^{\nu}g^{\mu \lambda}\right)\bigg \}  \ ,
\end{eqnarray}
where the $f_\phi^{\|}$ and $f_\phi^{\bot}$ are the decay constants of the $\phi$ meson \cite{Ball:2007zt,Ball:2007rt}, the $\zeta_{3\phi}$, $\kappa_{3\phi}$ and $\zeta_{4\phi}$, $\kappa_{4\phi}$ are  parameters in the twist-3 and twist-4 light-cone distribution amplitudes, respectively, which are defined in Ref.\cite{Ball:2007zt}.

Then we compute the integrals both in the coordinate space and in the momentum space, and obtain the correlation function $\Pi_{QCD}(p'^2,q^2)$, we introduce the subscript QCD to represent the QCD side.
The   $\Pi_{QCD}(p'^2,q^2)$ can be written as
\begin{eqnarray}
\Pi_{QCD}(p'^2,q^2)&=&\int_{4m_c^2}^{s^0_{J/\psi}}ds \frac{\rho_{QCD}(p'^2,s)}{s-q^2}+\cdots \ ,
\end{eqnarray}
through single-dispersion relation,
where the $\rho_{QCD}(p'^2,s)$ is the QCD spectral density,
\begin{eqnarray}
\rho_{QCD}(p'^2,s)&=& \lim_{\varepsilon \to 0}\frac{\text{Im} \Pi_{QCD}(p'^2,s+i\varepsilon) }{\pi}  \ ,
\end{eqnarray}
as the spectral density  $\rho_{QCD} (s',s)$ does not exist,
\begin{eqnarray}
\rho_{QCD} (s',s)&=&\lim_{\varepsilon' \to 0}  \lim_{\varepsilon \to 0}\frac{\text{Im}_{s'} \text{Im}_s \Pi_{QCD}(s'+i\varepsilon',s+i\varepsilon) }{\pi^2}  \nonumber\\
 &=&0 \ .
\end{eqnarray}
due to
\begin{eqnarray}
\lim_{\varepsilon' \to 0} \frac{\text{Im}_{s'} \Pi_{QCD}(s'+i\varepsilon',q^2) }{\pi}
 &=& 0 \ .
\end{eqnarray}

We rewrite the correlation function $\Pi_H (p'^2,q^2)$ on the hadron side as
\begin{eqnarray}
\Pi_H (p'^2,q^2)&=&\int_{4m_c^2}^{s^0_{X}}ds' \int_{4m_c^2}^{s^0_{J/\psi}}ds
\frac{\rho_H(s',s)}{(s'-p'^2)(s-q^2)}+\cdots \ ,
\end{eqnarray}
through double dispersion relation, where the $\rho_H(s',s)$  is the hadronic spectral density,
\begin{eqnarray}
\rho_H (s',s)&=&\lim_{\varepsilon' \to 0}  \lim_{\varepsilon \to 0}\frac{\text{Im}_{s'} \text{Im}_s \Pi_{H}(s'+i\varepsilon',s+i\varepsilon) }{\pi^2}  \ .
\end{eqnarray}

We math the hadron side  with the QCD side of the correlation function $\Pi(p'^2,q^2)$,
and carry out the integral over $ds'$  to obtain the rigorous duality \cite{Wang:2017lot,Wang:2019iaa},
\begin{eqnarray}
\int_{4m_c^2}^{s_{J/\psi}^0}ds \frac{\rho_{QCD}(p'^2,s)}{s-q^2}&=&\int_{4m_c^2}^{s_{J/\psi}^0} ds \frac{1}{s-q^2}\left[\int_{4m_c^2}^{\infty}ds'\frac{\rho_{H}(s',s)}{s'-p'^2}\right] \, ,\nonumber \\
&=&\left[\frac{\lambda_{X}m_{J/\psi}f_{J/\psi}g_{XJ/\psi\phi}}{
(m^2_{J/\psi}-q^2)(m^2_X-p'^2)}+\frac{C_{X'}}{m^2_{J/\psi}-q^2} \right] (m^2_X+m^2_{J/\psi}-m^2_\phi) \ .
\end{eqnarray}

The spectral densities $\rho_{H}(s',s)$ and $\rho_{QCD}(p'^2,s)$  are physical,
while the variables $p'^2$ and $q^2$ in Eq. (24) are free variables after performing the operator product expansion.
Generally speaking, we can set $p'^2=\alpha q^2$ with $\alpha$ to be a finite quantity.
According to the mass poles at $s'=m^2_X$ and $s=m_{J/\psi}^2$, we can obtain an approximated relation $s'=s$,
therefore, we set $p'^2=q^2$ and perform the  Borel transformation with respect to the variable $Q^2=-q^2$ to obtain the light-cone  QCD sum rules,
\begin{eqnarray}\label{gX}
&&\bigg\{\frac{\lambda_{X}m_{J/\psi}f_{J/\psi}g_{XJ/\psi\phi}}{
(m^2_X-m^2_{J/\psi})}\left[\exp\left(-\frac{m_{J/\psi}^2}{T^2}\right)
-\exp\left(-\frac{m_{X}^2}{T^2}\right)\right] \nonumber\\
&&+C_{X'}\exp\left(-\frac{m_{J/\psi}^2}{T^2}\right) \bigg\}
 \left(m^2_X+m^2_{J/\psi}-m^2_\phi \right) \nonumber\\
&=&\frac{\sqrt{2} i m_c}{2 \pi^2} f_\phi^\| m_\phi   \int_{4m_c^2}^{s_{J/\psi}^0} ds \int_{x_i}^{x_f} dx \, (1 - x) s \exp \left( -\frac{s}{T^2} \right)  \nonumber\\
&&-\frac{\sqrt{2}i}{8\pi^2} f_{\phi}^{\bot} m_{\phi}^2   \int_{4m_c^2}^{s_{J/\psi}^0} ds \int_{x_i}^{x_f} dx \, x(1 - x) \, (s - \tilde{m}_{c}^2) \exp \left( -\frac{s}{T^2} \right)  \nonumber\\
&&+\frac{\sqrt{2}i}{8\pi^2}  f_{\phi}^\bot m_{\phi}^2  \int_{4m_c^2}^{s_{J/\psi}^0} ds \int_{x_i}^{x_f} dx \, x (1 - x) \, [2 (s - \tilde{m}_{c}^2) + s] \exp \left( -\frac{s}{T^2} \right)      \nonumber\\
&&+\frac{\sqrt{2}i m_{c}^2}{8\pi^2}  f_{\phi}^\bot m_{\phi}^2   \int_{4m_c^2}^{s_{J/\psi}^0} ds \int_{x_i}^{x_f} dx \exp \left( -\frac{s}{T^2} \right)  \nonumber\\
&&-\frac{\sqrt{2}i m_{c}^3}{18T^2} \langle\frac{\alpha_s GG}{\pi} \rangle f_{\phi}^\| m_{\phi}  \int_{0}^{1} dx \frac{1}{(1-x)^2} \left( 1-\frac{\tilde{m}_c^2}{2T^2} \right) \exp \left( -\frac{\tilde{m}_c^2}{T^2} \right)
 \nonumber\\
&&-\frac{\sqrt{2}i m_{c}^3}{18T^2} \langle\frac{\alpha_s GG}{\pi} \rangle f_{\phi}^\| m_{\phi}  \int_{0}^{1} dx \frac{(1-x)}{x^3} \left( 1-\frac{\tilde{m}_c^2}{2T^2} \right) \exp \left( -\frac{\tilde{m}_c^2}{T^2} \right)
 \nonumber\\
&&+\frac{\sqrt{2}i m_{c}}{12} \langle\frac{\alpha_s GG}{\pi} \rangle f_{\phi}^\| m_{\phi}  \int_{0}^{1} dx \frac{(1-x)}{x^2} \left( 1-\frac{\tilde{m}_c^2}{T^2} \right) \exp \left( -\frac{\tilde{m}_c^2}{T^2} \right)
 \nonumber\\
&&-\frac{\sqrt{2}i m_{c}^2}{72T^2} \langle\frac{\alpha_s GG}{\pi} \rangle f_{\phi}^\bot m_{\phi}^2  \int_{0}^{1} dx
\frac{x}{(1-x)^2} \left( 1-\frac{\tilde{m}_c^2}{T^2} \right) \exp \left( -\frac{\tilde{m}_c^2}{T^2} \right)
 \nonumber\\
&&+\frac{\sqrt{2}i m_{c}^4}{72T^4} \langle\frac{\alpha_s GG}{\pi} \rangle f_{\phi}^\bot m_{\phi}^2  \int_{0}^{1} dx
\frac{1}{(1-x)^3}   \exp \left( -\frac{\tilde{m}_c^2}{T^2} \right)
 \nonumber\\
&&-\frac{\sqrt{2}i m_{c}^2}{24T^2} \langle\frac{\alpha_s GG}{\pi} \rangle f_{\phi}^\bot m_{\phi}^2  \int_{0}^{1} dx
\frac{1}{(1-x)^2} \exp \left( -\frac{\tilde{m}_c^2}{T^2} \right)
 \nonumber\\
&&+\frac{\sqrt{2}i m_{c}}{36} \langle\frac{\alpha_s GG}{\pi} \rangle f_{\phi}^\| m_{\phi}  \int_{0}^{1} dx
\frac{1}{x} \left( 1-\frac{\tilde{m}_c^2}{T^2} \right) \exp \left( -\frac{\tilde{m}_c^2}{T^2} \right)
 \nonumber\\
&&-\frac{\sqrt{2}i}{144} \langle\frac{\alpha_s GG}{\pi} \rangle f_{\phi}^\bot m_{\phi}^2  \int_{0}^{1} dx
 \left( 2+\frac{\tilde{m}_c^2}{T^2} \right) \exp \left( -\frac{\tilde{m}_c^2}{T^2} \right)
 \nonumber\\
&&+\frac{\sqrt{2}i m_c^2}{8\pi^2}  f_{\phi}^\bot m_{\phi}^4 \zeta_{4\phi}^\bot
 \int_{0}^{1} dx
\frac{1}{(1-x)}  \exp \left( -\frac{\tilde{m}_c^2}{T^2} \right)
 \nonumber\\
&&+\frac{\sqrt{2}i m_c}{8\pi^2}  f_{\phi}^\| m_{\phi}^3 \zeta_{4\phi}^\|
 \int_{0}^{1} dx
\frac{(1-x)}{x}\tilde{m}_c^2  \exp \left( -\frac{\tilde{m}_c^2}{T^2} \right)
 \ ,
\end{eqnarray}
where $x_{f}=\frac{1+\sqrt{1-4m_{c}^{2}/s}}{2}$, $x_{i}=\frac{1-\sqrt{1-4m_{c}^{2}/s}}{2}$,  $\tilde{m}_{c}^{2}=\frac{m_{c}^{2}}{x(1-x)}$,
and the $T^2$ is the Borel parameter.
In numerical calculations, we take the $C_{X'}$ as a free parameter,
and choose the suitable value  to obtain the stable QCD sum rules with the variation of the $T^2$.

\section{Numerical results and discussions}
At the QCD side,
we take the standard value of the vacuum condensate $\langle\frac{\alpha_{s}GG}{\pi}\rangle=(0.012\pm0.004\,\text{GeV}^{4})$
\cite{Shifman:1978bx,Shifman:1978by,Reinders:1984sr,Colangelo:2000dp},
and take the $\overline{MS}$ mass $m_{c}(m_{c})=(1.275\pm0.025)\,\text{GeV}$ from the Particle Data Group \cite{ParticleDataGroup:2024cfk}.
Moreover, we take into
account the energy-scale dependence of the $\overline{MS}$ mass from the renormalization group equation,
\begin{eqnarray}
m_{c}(\mu)&=&m_{c}(m_{c})\bigg[\frac{\alpha_{s}(\mu)}{\alpha_{s}(m_{c})}\bigg]^{\frac{12}{33-2n_f}}\ ,\nonumber\\
\alpha_{s}(\mu)&=&\frac{1}{b_{0}t}\left[1-\frac{b_{1}}{b_{0}^{2}}\frac{\log t}{t}+\frac{b_{1}^{2}(\log^{2}t-\log t-1)+b_{0}b_{2}}{b_{0}^{4}t^{2}}\right]\ ,
\end{eqnarray}
where $t=\log \frac{\mu^{2}}{\Lambda^{2}}$, $b_{0}=\frac{33-2n_{f}}{12\pi}$, $b_{1}=\frac{153-19n_{f}}{24\pi^{2}}$, $b_{2}=\frac{2857-\frac{5033}{9}n_{f}+\frac{325}{27}n_{f}^{2}}{128\pi^{3}}$, $\Lambda=210\,\text{MeV}$, $292\,\text{MeV}$ and $332\,\text{MeV}$
for the flavors $n_{f}=5$, $4$ and $3$, respectively \cite{ParticleDataGroup:2024cfk,Narison:1983kn},
and we choose the flavor number to be $n_f=4$ because of the concerning of the $c$-quark, and evolve the input parameters to the typical energy scale  $\mu=1.275\,\text{GeV}$.
In addition, the parameters $f_\phi^\|$, $f_\phi^\bot$, $\zeta_{4\phi}^\bot$ and $\zeta_{4\phi}^\|$  are taken as  $f_\phi^\|=0.215\,\text{GeV}$,
$f_\phi^\bot=0.186\,\text{GeV}$, $\zeta_{4\phi}^\bot=-0.01$  and  $\zeta_{4\phi}^\|=0$ \cite{Ball:2007zt}.

At the hadron side,
we take the  $m_{J/\psi}=3.0969\,\text{GeV}$, $m_{\phi}=1.01946\,\text{GeV}$ from the Particle Data Group \cite{ParticleDataGroup:2024cfk},
$\sqrt{s_{J/\psi}^0}=3.6\, \text{GeV}$, $f_{J/\psi}=0.418\,\text{GeV}$ \cite{Becirevic:2013bsa},  $m_{X}=4.1465\,\text{GeV}$ \cite{ParticleDataGroup:2024cfk}, $\lambda_{X}=2.88 \times 10^{-2}\,\text{GeV}^5$  from the QCD sum rules \cite{Wang:2024ciy}.

\begin{figure}[htp]
\centering
\includegraphics[totalheight=6cm,width=9cm]{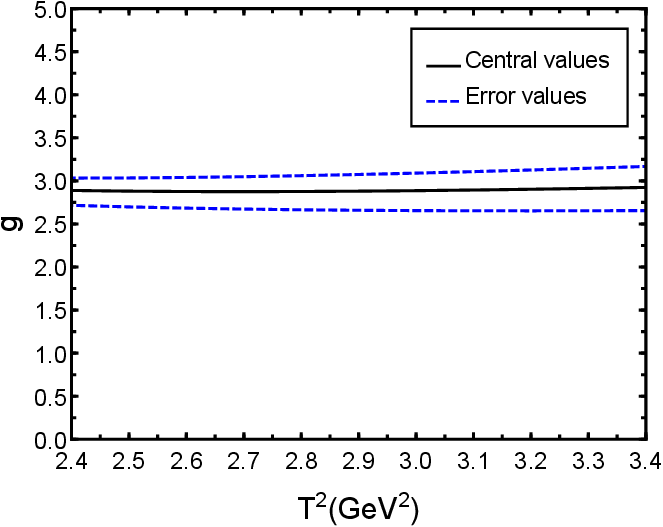}
\caption{The hadronic coupling constant $g_{XJ/\psi \phi}$ with variation of the Borel parameter $T^2$.}\label{fig:fig10}
\end{figure}

In numerical calculation, we fit the free parameter to be $C_{X'}=0.000655(T^2-2\, \text{GeV}^2)\,\text{GeV}^5$
to obtain the platform in the Borel window  $T^2=(2.4-3.4)\,\text{GeV}^2$, the interval region of the Borel parameter between the maximum and  minimum values is about $1\,\rm{GeV}^2$, just like in our previous works \cite{Wang:2017lot,Wang:2019iaa,
WZG-Y4500-decay,WZG-Zcs4123-decay,WZG-Zcs3985-decay,
WZG-Y4500-NPB-2024,WZG-YXS-X6552}.
In Fig.\ref{fig:fig10}, we plot the hadronic coupling constant $g_{X J/\psi \phi}$  in regard to variation of the Borel parameter.
In the Borel window, there appears very flat platform indeed, it is reasonable and reliable to extract the hadron coupling constant.
After taking into account the uncertainties of the input parameters, we obtain the value of the hadronic coupling constant $g_{X J/\psi \phi}$,
\begin{eqnarray}
g_{X J/\psi \phi}&=&2.88 \pm 0.21 \, .
\end{eqnarray}
Now it is easy to obtain the decay width,
\begin{eqnarray}
\Gamma_{X(4140)\rightarrow J/\psi \phi}&=&\frac{p(m_X,m_{J/\psi},m_\phi)}{24\pi m_X^2} g_{XJ/\psi\phi}^2 \bigg\{ \frac{\left( m_X^2-m_\phi^2\right)^2}{2m_{J/\psi}^2}
+\frac{\left( m_X^2-m_{J/\psi}^2\right)^2}{2m_{\phi}^2}
 \nonumber\\
&&+4m_X^2-\frac{m_{J/\psi}^2+m_{\phi}^2}{2} \bigg\} \, , \nonumber\\
&=&145 \pm 21 \,\text{MeV}\, .
\end{eqnarray}
where $p(a,b,c)=\frac{\sqrt{[a^2-(b+c)^2][a^2-(b-c)^2]}}{2a}$.
The width $\Gamma_{X(4140)\rightarrow J/\psi \phi}=145 \pm 21 \,\text{MeV}$ is reasonable according to the measured data
$162\pm21^{+24}_{-49 } \text{MeV}$ by the LHCb collaboration in 2021.
We obtain additional support that the predictions from the QCD sum rules favors assigning the $X(4140)$ as the $[sc]_S[\bar{s}\bar{c}]_A+[sc]_A[\bar{s}\bar{c}]_S$ type axialvector tetraquark state with $J^{PC} = 1^{++}$. We can extend present work to study the strong decays of other tetraquark candidates with the light-cone QCD sum rules to diagnose their nature.

\section{Conclusion}
In this article,
we take the $X(4140)$ as the 1S $[sc]_S[\bar{s}\bar{c}]_A+[sc]_A[\bar{s}\bar{c}]_S$ tetraquark state with the $J^{PC} = 1^{++}$ to study the hadronic coupling constant $g_{XJ/\psi\phi}$ via the light-cone QCD sum rules, then calculate the width of the two-body strong decay $X(4140)\rightarrow J/\psi \phi$.
In calculations, we introduce free parameters to parameterize the  contributions of the higher resonances and continuum states,
resort to the rigorous quark-hadron duality to match the hadron side with the QCD side of the correlation function,
 then obtain the light-cone QCD sum rules for the hadronic coupling constant.
Through varying the free parameter,
we obtain the flat Borel platform and extract the value of the hadronic coupling constant. 
The corresponding decay width $\Gamma(X(4140)\rightarrow J/\psi \phi)=145\pm21 \,\text{MeV}$  is reasonable according to the updated measurement $162\pm21^{+24}_{-49 }\, \text{MeV}$  by the LHCb collaboration.
The numerical results lead to additional support for the possibility that $X(4140)$ could be the $[sc]_S[\bar{s}\bar{c}]_A+[sc]_A[\bar{s}\bar{c}]_S$ type axialvector tetraquark state, as we cannot assign a hadron unambiguously with the mass alone.

\section*{Acknowledgements}
This  work is supported by National Natural Science Foundation, Grant Number
12575083.

\end{document}